\documentclass[11pt]{article}
\usepackage[dvips]{graphics}

\begin{document}

\noindent
\Large
{\bf CAUSAL PARADOXES: A CONFLICT BETWEEN \\
RELATIVITY AND THE ARROW OF TIME}
\normalsize
\vspace*{1cm}

\noindent
{\bf Hrvoje Nikoli\'c}

\vspace*{0.5cm}
\noindent
{\it
Theoretical Physics Division \\
Rudjer Bo\v{s}kovi\'{c} Institute \\
P.O.B. 180, HR-10002 Zagreb, Croatia \\
E-mail: hrvoje@thphys.irb.hr}

\vspace*{2cm}

\noindent
It is often argued that superluminal velocities and nontrivial
spacetime topologies, allowed by the theory of relativity, may
lead to causal paradoxes. By emphasizing that the notion of
causality assumes the existence of a time arrow (TA) that points from
the past to the future, the apparent paradoxes appear to be
an artefact of the wrong tacit assumption that the relativistic
coordinate TA coincides with the physical TA.
The latter should be identified with the thermodynamic TA,
which, by being absolute and irrotational, does not lead to paradoxes.
\vspace*{0.5cm}


\section{INTRODUCTION - A BRIEF OVERVIEW OF CAUSAL PARADOXES}

The principle of causality says that the past influences the future, 
while the future does not influence the past. Clearly, this 
principle assumes the existence of a direction of time, that is, 
a {\em time arrow} (TA) that points from the past to the future. 
However, according to the theory of relativity, time is not 
absolute. In particular, the time ordering of points in spacetime 
depends on the choice of the time coordinate. It is often argued 
that this relativity of time may lead to causal paradoxes, provided 
that superluminal propagation of information is possible 
\cite{tolold,bil,ben,rec,dol}, 
or that the geometry and topology of spacetime allows closed timelike 
curves \cite{god,tip,mor,vis}. 
The superluminal propagation 
of information may be realized through the propagation of particles 
with negative squared mass (tachyons) \cite{bil1,bil,rec}, 
or through a modification of the dispersion relation 
(induced by interactions) of massless particles
\cite{gar,chu,drum,sch,chi,bol,chi2,nik}.
A spacetime geometry and topology that allows closed timelike 
curves may be realized with wormholes that require negative energy 
\cite{mor1,mor,vis2} and with other mechanisms 
\cite{god,tip,gott}. There is a lot of work trying to prove 
the so called {\it chronology protection conjecture} 
\cite{hawk,des,grant,sus,vis,lib},
which essentially says that that the laws of physics conspire 
so that the  physical conditions needed for a causal paradox 
cannot realize in nature. However, a convincing proof of this 
conjecture is still far from being accomplished
\cite{kim,tan,kras,vis,visPRL,kuh}.  

The basic idea for a construction of a causal paradox is allways 
the same. One has to construct a causal loop, that is, a 
{\em globally oriented} closed curve in spacetime, such that the orientation 
corresponds to the direction from the causes to the consequences.
The principle of causality says that the causes influence the 
consequences, while the consequences do not influence the causes.
However, since the curve is closed, it is necessary that a cause 
eventually becomes a consequence of its own consequence, which 
may lead to a causal paradox. From the relativistic point of view,
the paradox can be viewed in the following way. 
Locally, each point of the causal loop is oriented 
from the past to the future, as seen by the corresponding 
local observer. (When the curve is timelike, the ``corresponding" 
observer may be identified with an observer that moves along this curve. 
When the curve is locally lightlike or spacelike, the ``corresponding" 
observer may be identified with an observer for which 
this part of the curve 
is oriented from the past to the future.) However,  
in a fixed {\em global}
system of coordinates, some parts of the causal loop 
may be oriented from the future to the past, while others may be 
oriented 
from the past to the future. This means that, along some parts of 
the curve, the future may influence the past, which contradicts 
the principle of causality. 
The aim of this paper is to present a general solution of the 
causal paradoxes that may appear within the scheme above. 

Before starting with a detailed analysis, let us briefly present  
the conceptual picture for which we present 
arguments in the subsequent sections of the paper. At the fundamental 
level, time should be treated on an equal footing with space. 
In this sense, there can be no time-travel paradoxes, 
just as there are no space-travel paradoxes. Time does not lapse, 
just as space does not lapse. Not every initial condition needs 
to lead to a consistent solution (of an equation of motion), 
just as not every boundary 
condition needs to lead to a consistent solution. This is not 
inconsistent as long as there exist consistent global solutions 
that are not 
necessarily found by fixing an initial or a boundary condition.
The notion of causality 
does not make sense at the fundamental level. Without causality, 
there can be no causal paradoxes. However, in order to explain why 
many people think that there are causal paradoxes, we need to explain 
why they think that there is causality, or why they think that time lapses, 
or why they think that 
every initial condition should lead to a consistent solution.     
We find that the origin of all such time-space asymmetric thinkings
can be traced back to the existence of the thermodynamic TA, 
which, in turn, is not a fundamental phenomenon. In other words, since people 
live in a universe with a thermodynamic TA, it seems to them
that there are possibilities for causal paradoxes, 
despite the fact that in reality there are 
no possibilities for causal paradoxes at all.   

\section{THE ORIGIN OF THE ARROW OF TIME}

The key point of the solution is the fact that the fundamental 
laws of physics (at least those that we know)
do not contain any preferred direction of time. 
There is no difference between the future and the past at the 
fundamental level. If the Cauchy problem corresponding 
to the fundamental equations of motion is well posed, then the initial 
data on a Cauchy surface determine the physical quantities 
in the {\em both} directions from the Cauchy surface. Therefore, 
from the fundamental point of view, it is meaningless to say 
that the past influences the future, but that the future does 
not influence the past. In this sense, the principle of causality 
formulated as above is not a fundamental physical principle.
 
The principle of causality is only an effective phenomenological 
physical law, valid only on the macroscopic level. The TA 
pointing from the past to the future exists only on the macroscopic 
level. There are various phenomenological manifestations of the 
TA, such as the causal TA, the psychological TA  
and the electrodynamic TA.
However, they all can be reduced to the thermodynamic TA, that is, 
to the phenomenological rule that entropy (or some other measure 
of disorder) increases with time 
\cite{feyn,dav,zeh,leb}. 
(For popular explanations, see also Refs. \cite{feyn2,hawkbook}.) 
Therefore, {\em the physical TA is the thermodynamic TA.}

In particular, the origin 
of the causal TA can be qualitatively understood as follows. 
When the Cauchy problem is well posed, then the complete knowledge of 
the present state 
determines uniquelly and self-consistently both the future and the 
past. However, in practice, we try to draw conclusions about 
the future and the past without knowing {\em details} of the present 
state. Thus, our conclusions are to a great extent based on 
statistical arguments. Since disorder increases with time, 
statistical arguments can determine the past much better than 
the future. Therefore, it is typically easier to see a relation 
between the presence and the past than to see a relation between 
the presence and the future. This is why, in practice,
we consider the past 
as a cause of the presence, but we do not consider the future 
as a cause of the presence. 

The causal TA is also closely related to the psychological TA, 
that is, to our psychological experience that time lapses 
from the past to the future. This experience is a consequence of the 
rule that brains remember the past, but not the future. As recalling 
allways reduces to an observation of a present state 
(of the medium that stores the data) and 
subsequent conclusion about events that do not refer to the 
present time, it is clear that the psychological TA reduces 
to the causal TA. (For a recent detailed discussion of the 
psychological TA see Ref. \cite{hart}.)

%
%

The fact that we live in a universe in which entropy 
increases with time means 
that, in the past, the universe was in a state with a very low 
entropy. Such a low entropy state is very improbable and we do 
not know why the nature have chosen such improbable initial 
conditions. (For a recent attempt to explain this, see 
Refs. \cite{carr,nikcom}.
For a general discussion, see, e.g., Ref. \cite{price}.)
However, the point is that the existence of the 
thermodynamic TA is merely a macroscopic property 
of the specific solution in which we live. 
(In fact, even in an equilibrium where a TA does not exist,
the thermodynamic state still defines the physical time
in a fully general-relativistic background-independent 
manner \cite{rov1,rov2}.)

\section{RELATIVISTIC DEFINITION OF THE THERMODYNAMIC TA}

In order to give a more quantitative discussion,
let us introduce the entropy density $s(x)$ \cite{tol,nikolNPB}, defined
everywhere in the universe. It transforms as a scalar 
under general coordinate transformations. (From this scalar, 
one can also construct the entropy vector $s^{\mu}=s\, dx^{\mu}/d\tau$,
where $dx^{\mu}/d\tau$ is the macroscopic 
local 4-velocity of a fluid \cite{tol}.
Alternatively, if one starts from the entropy vector as a more
fundamenatl quantity, one can define the entropy scalar as
$s=|s^{\mu}s_{\mu}|^{1/2}$.)   
We can define the local TA in an absolute 
and relativistically covariant way, 
as a direction of the gradient of $s$
\begin{equation}\label{grad}
t_{\mu}=\partial_{\mu}s.
\end{equation}
In fact, the vector $t_{\mu}$ does not need to be timelike, but 
we live in a universe in which it is usually timelike, 
so we retain the name ``time arrow" (TA). 
Along a given curve in spacetime, one can decompose $t_{\mu}$ as
\begin{equation}\label{tang}
t_{\mu}=t_{\mu}^{\parallel}+t_{\mu}^{\perp}, 
\end{equation}
where $t_{\mu}^{\parallel}$ is the component of $t_{\mu}$ tangential 
to the curve, while $t_{\mu}^{\perp}$ is the component normal to the 
curve. The vector $t_{\mu}^{\parallel}$ defines the local orientation 
of the curve. The physical TA on the curve is given by 
$t_{\mu}^{\parallel}$. If the curve is a timelike trajectory 
of a local observer, then the TA locally seen by this observer is given 
by $t_{\mu}^{\parallel}$, which is a timelike vector.
From (\ref{tang}) and (\ref{grad}) we see 
that the integral of $t_{\mu}^{\parallel}$ over any closed curve is
\begin{equation}\label{int}
\oint dx^{\mu}t_{\mu}^{\parallel}=\oint dx^{\mu}t_{\mu}=0.
\end{equation}
This implies that a closed curve cannot possess 
a {\em physical} global orientation. Consequently, a  
causal loop cannot exist, which eliminates a possibility 
for a causal paradox. It is a simple consequence of the fact 
that the physical TA is a gradient, which implies that it is 
irrotational.

\section{RELATION BETWEEN CAUSAL PARADOXES AND THE THERMODYNAMIC TA}

Now it is easy to understand the source of apparent causal 
paradoxes. Consider, for example, a closed timelike curve
$x^{\mu}(\tau)$, where $\tau$ is an affine parameter 
on the curve equal to the corresponding relativistic 
proper length. The orientation of the curve can be chosen 
to be directed from a smaller to a larger value of $\tau$. 
The proper length can be interpreted 
as the proper time of the observer whose trajectory 
coincides with the curve. Assuming that such an orientation
correponds to the physical direction of the proper time, 
one finds a causal paradox. However, such an orientation 
is only a coordinate orientation. The true physical orientation 
is defined locally by $t_{\mu}^{\parallel}$. It is impossible that
the coordinate orientation above everywhere coincides 
with the physical orientation. Instead, at 
some parts of the closed curve, the coordinate TA has the 
direction opposite to the physical TA.  

The explanation above requires some additional clarifications.
One may object that there are realistic systems in which 
the local entropy density {\em decreases} with the physical time, 
which is consistent with the thermodynamical laws which say that 
only the total entropy of a large isolated system
must increase with the physical time. We have two responses to such 
objections.
First, in such systems, the 
physical time is defined globally, as a coordinate with respect to which
the total entropy increases. An observer which only measures the local 
entropy will naturally assign a different orientation of physical time.
(After all, we cannot be sure that in some distant parts of the universe
invisible by us the entropy of isolated systems 
does not decrease with the usual globally defined coordinate time.) 
Second, for observers that are able to measure entropy in large 
regions of spacetime, it is natural to divide the spacetime into 
appropriately large regions and to assign the entropy to each region
as the average entropy of this region. This division of spacetime induces 
also the division of any curve in spacetime. In general, the neighboring 
regions of a curve have different entropies, which defines the local 
orientation of a curve on each boundary between two regions on the curve.
Again, it is easy to see that such a procedure cannot assign a global 
orientation to a closed curve, because, if not all regions have the 
same entropy, there will be at least two boundaries that have relative 
orientations that are inconsistent with the global orientation.   
This again solves the causal paradoxes.

Some forms of the causal paradox rest on the fact that, for some 
spacetime topologies, the Cauchy problem is not well posed 
\cite{fried}. The simplest example is a cylindrical spacetime 
in which the time coordinates $t=0$ and $t=2\pi$ are identified. 
As a solution $\phi(t,{\bf x})$ of the equations of motion must 
be periodic such that $\phi(t,{\bf x})=\phi(t+2\pi,{\bf x})$, it is 
clear that not every initial condition leads to a solution. 
If one assumes that, in principle, any initial condition 
can be realized in nature, 
then one may obtain a paradox consisting in finding a solution 
for which $\phi(t,{\bf x}) \neq \phi(t+2\pi,{\bf x})$. At first 
sight, such forms of the paradoxes have nothing to do with 
the thermodynamic TA. However, this is not completely true; 
there is an indirect relation with the thermodynamic TA. 
Mathematically, the source of the paradox lies in the wrong 
assumption that any initial condition should lead to a solution.
Indeed, if one abandones this wrong assumption, then there 
is no paradox at all. But the fact that some people think
of it as a paradox indicates that there exists some deeper 
reason for retaining this assumption. To identify this reason, 
consider a cylindrical spacetime in which it is the {\em space} 
coordinate that is compactified. Now it is the boundary 
condition (not the initial condition) that cannot be arbitrary. 
Physicists find nothing paradoxical with having a constraint 
on the boundary condition. So why 
they find a paradox when it is the {\em initial} condition that
is constrained, so that it 
cannot be arbitrary? Obviously, contrary to the 
fundamental principle of 
relativity, they do not treat time on the equal footing with 
space. But the only physical source of different  
treatments of space and time is the existence of the physical TA, 
which, of course, is the thermodynamic TA.  
   
For a complete discussion of causal paradoxes, it is
unavoidable to mention the relation with the concept 
of ``free will" \cite{rec,vis,fried}.
Some physicists would like to retain the concept of ``free will"
compatible with physical laws. They often think of 
``free will" as something that determines the initial conditions, 
so they view a constraint on an initial condition as a 
constraint on ``free will", which they often find unsatisfying.
However, the usual notion of ``free will" that influences the future 
but not the past is clearly related to the psychological 
TA, which, in turn, is a consequence of the thermodynamic TA.  
A notion of a ``fundamental free will" compatible with 
the fundamental physical laws should treat time on the equal 
footing with space, so there is no any 
{\it a priori} reason why ``free will" 
should act in terms of initial conditions.  

At the fundamental relativistic level, time should be treated 
on the equal footing with space. 
All physical quantities are described by certain 
functions $\phi(x)$, where $x$ are points on spacetime. 
The functions $\phi(x)$ are required to be single valued. (In fact, 
if they are not single valued, then they are not functions at all.)
Clearly, the requirement of single valuedness automatically 
excludes paradoxes. This requirement is sometimes postulated
\cite{nov} or derived from other physical laws \cite{car1,car2} 
as the principle of self-consistency, 
which serves as a general mechanism that prevents causal 
paradoxes. In our view, the principle of self-consistency 
is merely a tautology that does not need a separate postulate 
or an additional justification (see also Ref. \cite{deut}). 
It needs a separate postulate 
or a justification if one starts from the assumption that 
the universe is fundamentally determined by its {\em initial} condition. 
However, such an assumption is not relativistic in spirit, but 
is related to our subjective experience of time that reflects 
the existence of the thermodynamic TA. Therefore, we abandon 
this assumption.

\section{CONCLUSION}

Our results can be summarized as follows. 
The apparent causal paradoxes are an artefact of the wrong 
tacit assumption that the relativity of time implies also 
the relativity of the time arrow. Instead, the TA 
that makes the distinction between the past and the future is absolute,
but defined only on the macroscopic phenomenological level.
Since the theory of relativity does not allow the existence 
of an absolute preferred time direction,
this constitutes an apparent conflict between relativity 
and the arrow of time. However, there is no real conflict, 
because the TA is merely a property of a specific solution 
of (presumably relativistic) equations of motion that determine 
our universe. 

\vspace{0.4cm}
\noindent
{\bf Acknowledgements.}
The author is grateful to K.~Kumeri\v{c}ki for his 
critical reading of the manuscript and useful remarks.
This work was supported by the Ministry of Science and Technology of the
Republic of Croatia under Contract No.~0098002.

\end{document}